\documentclass{mem}
\usepackage{natbib}\usepackage{txfonts}\usepackage{balance}
\usepackage{graphicx}
\usepackage[a4paper,breaklinks,dvipdfm]{hyperref}
\idline{75}{282}
\begin{document}
\def\teff{$T\rm_{eff }$}
\def\kms{$\mathrm {km s}^{-1}$}

\title{
Chemical element abundances in the outer halo globular cluster M~75}

   \subtitle{}

\author{
N. \,Kacharov
\and A. \, Koch
          }

  \offprints{N. Kacharov}

\institute{Landessternwarte, Zentrum f\"{u}r Astronomie der Universit\"{a}t Heidelberg, K\"{o}nigstuhl 12, D-69117 Heidelberg, Germany
\email{n.kacharov@lsw.uni-heidelberg.de}
}

\authorrunning{Kacharov}

\titlerunning{Chemical element abundances in M~75}

\abstract{We present the first comprehensive abundance study of the massive, outer halo globular cluster (GC) M~75 (NGC~6864). This unique system shows a very extended trimodal horizontal branch (HB), but no other clues for multiple populations have been detected in its colour-magnitude diagram (CMD). Based on high-resolution spectroscopic observations of 16 red giant stars, we derived the abundances of a large variety of $\alpha$, p-capture, iron-peak, and n-capture elements. We found that the cluster is metal-rich ([Fe/H]~$= -1.16 \pm 0.02$~dex, [$\alpha$/Fe]~$= +0.30 \pm 0.02$~dex), and shows a marginal spread in [Fe/H] of $0.07$~dex, typical of most GCs of similar luminosity. We detected significant variations of O, Na, and Al among our sample, suggesting three different populations. Additionally, the two most Na-rich stars are also significantly Ba-enhanced, indicating a fourth population of stars. Curiously, most stars in M 75 (excluding the two Ba-rich stars) show a predominant r-process enrichment 
pattern, which is unusual at the cluster's high metallicity. We compare the abundance properties of M 75 and NGC 1851 (a GC very similar to M 75 in terms of age, metallicity, and HB morphology) and draw conclusions on M 75's possible formation scenarios.
\keywords{Stars: abundances --
Globular clusters: individual: M~75 -- 
Galaxy: halo --}
}
\maketitle{}

\section{Introduction}

\begin{figure*}[t!]
\centering
\includegraphics[height=.34\textheight]{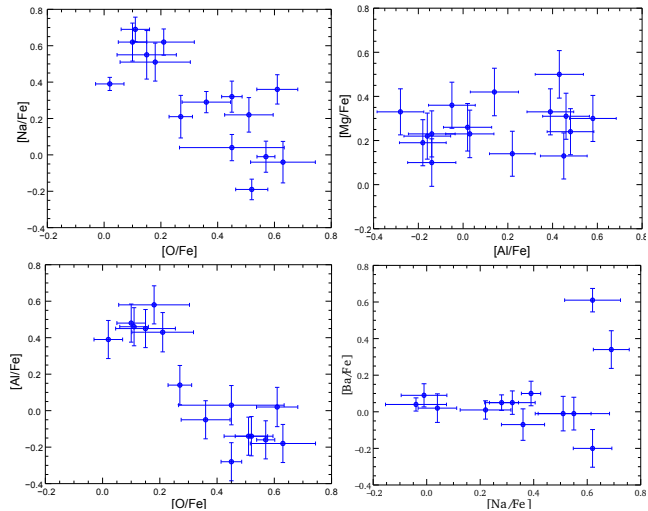}
\caption{\footnotesize
Abundance variations in M 75: O-Na and O-Al anticorrelations, (no) Al-Mg anticorrelation, and Ba-rich stars.
}
\label{anticor}
\end{figure*}

With ages of 10 -- 14 Gyr, Globular GCs are amongst the oldest stellar systems in the Milky Way. Long time considered as simple stellar populations, nowadays we recognise their complex star formation histories through precise abundance analyses of a variety of chemical elements in individual cluster members. Although we do not necessarily see clues for multiple populations in all GC CMDs, all of them studied to date present significant spreads and certain anticorrelations between their light and $\alpha$-element abundances \citep{gratton+2012}. These are tightly linked to the ``second parameter'' effect, which needs to explain discordant Horizontal Branch (HB) morphologies at any given metallicity \citep{dantona+2002}. CNO or He-content variations amongst the stars in a GC are responsible for the appearance of very extended HBs and are accompanied by variations of the p-capture elements \citep{gratton+2011,gratton+2012b}.

We obtained high-resolution (R~$\sim 30 000$) spectra of 16 bright giant stars in M~75 with the MIKE spectrograph mounted at the 6.5-m Magellan telescope aiming at a comprehensive abundance analysis of this GC. It is located at a Galactocentric distance of $15$~kpc, which tenants the transition region between the inner and outer Milky Way halo. Its younger age \citep[$\sim10$~Gyr;][]{catelan+2002} and high metallicity ([Fe/H]$=-1.16$~dex) are compatible with the properties of the outer halo GC system and suggest a possible extragalactic origin. On the other hand, M~75 is amongst the most concentrated GCs ($c=\log(r_t/r_c)=1.80$), which could be contrasted to the extended and loose clusters in the outer halo \citep{koch+cote2010,koch+2009}. This unique GC also has a trimodal HB, which is not explicable under canonical stellar evolutionary models  \citep{catelan+2002}. Apart from the well separated red HB (RHB) and blue HB (BHB), its CMD shows a distinct third extension of a very blue, faint tail. Thus, it is 
very important to assess possible multiple populations, which could be related to the peculiar HB morphology and to look for peculiarities in its chemical composition, which might reveal clues about its origin and early evolution.

\section{Light elements in M~75 -- clues for multiple populations}

We detected significant variations in the star-to-star abundances of Na, O, and Al in M 75 but not in Mg, which would be typical for bright GCs (Fig. \ref{anticor}). The stars with [Na/Fe]~$ < 0.1$~dex are considered to be the remainder of the primordial first generation (FG) of stars. They are similar to the stars in the Galactic halo. In contrast, stars from the second generation (SG) show considerably higher Na- and lower O-abundances but they are indistinct in terms of other chemical elements. We also found that the two most Na-rich stars are Ba-enhanced by $0.4$ and $0.6$~dex, respectively, above the cluster's mean solar [Ba/Fe] ratio, indicative of s-process enrichment from AGB stars.

\begin{figure}[t!]
\resizebox{\hsize}{!}{\includegraphics[clip=true]{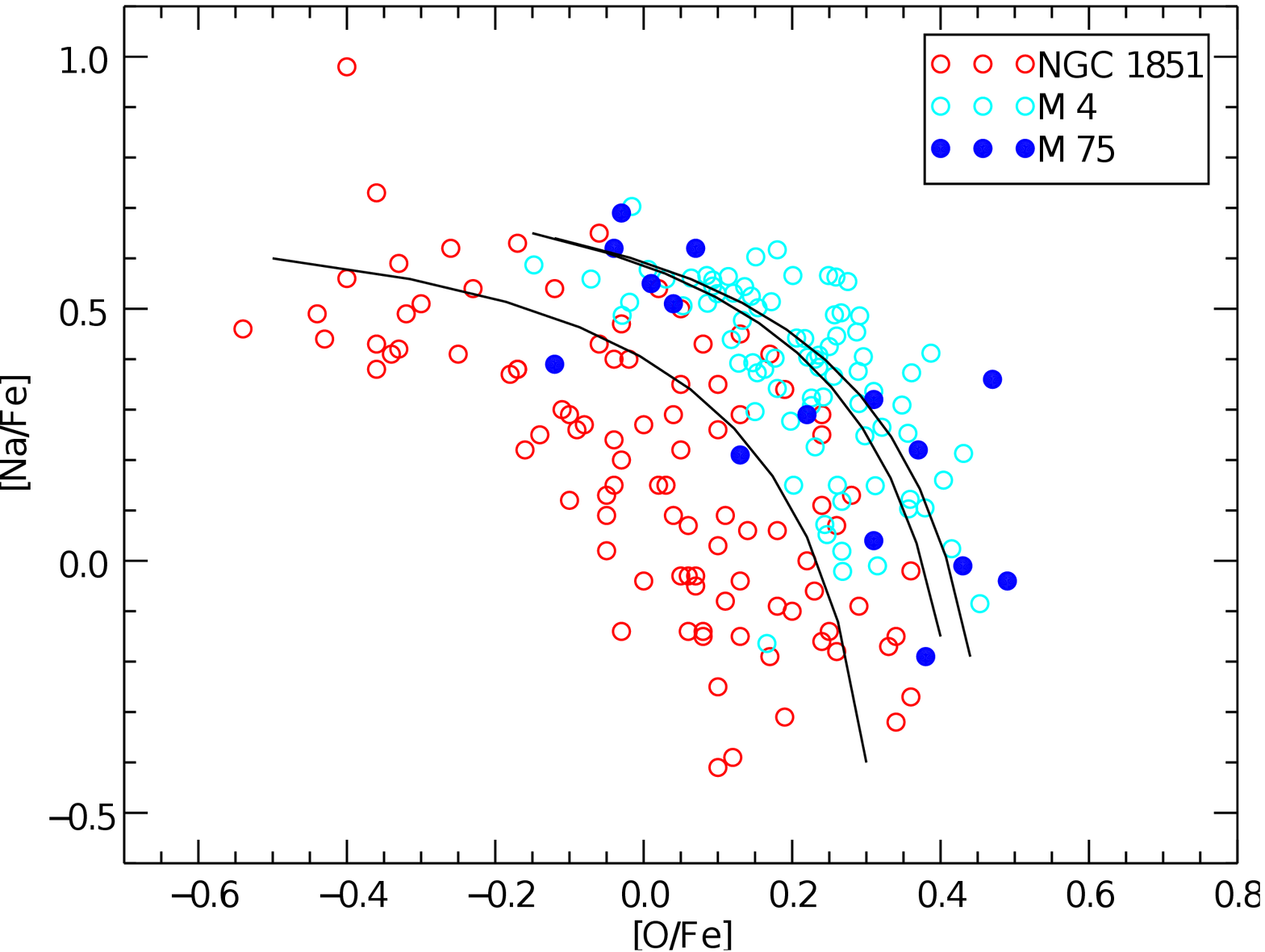}}
\caption{\footnotesize
The Na-O anticorrelations in M~75, M~4 \citep{carretta+2009} and NGC~1851 \citep{carretta+2011}. Simple dilution models \citep{carretta+2009} for each GC are overimposed.
}
\label{anticor2}
\end{figure}

The extended blue tail of M 75's HB is at odds with our findings of only a moderate Na-O anticorrelation, which so far lacks an extreme population on the RGB, characterised by extremely O-poor stars. The Na-O anticorrelation of M 75 better resembles the less massive GC M~4 than that of NGC~1851 (Fig. \ref{anticor2}). The latter GC is often thought as M 75's twin in terms of age, luminosity, and HB morphology. However, its CMD shows double RGB and SGB, which are not found in M 75. The less extended Na-O anticorrelation of M 75 suggests lower mass AGB polluters \citep{carretta+2009}, which is further supported by the presence of Ba-enhanced stars in this cluster. 

The Na-O anticorrelation (although not very extended) is consistent with three different stellar populations in M~75. Complemented with the two Ba-rich stars, we suggest the presence of four chemically distinct populations in this GC (Fig. \ref{anticor}).


\section{n-capture elements in M~75 -- s-process deficient primordial stars}

\begin{figure}[t!]
\resizebox{\hsize}{!}{\includegraphics[clip=true]{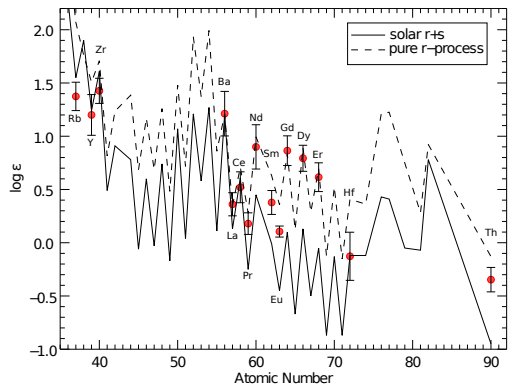}}
\caption{\footnotesize
Mean neutron capture elements measurements in M 75, normalised to Ba. The lines display the solar and r- and s-process contributions from \citet{burris+2000}. 
}
\label{rs}
\end{figure}

M~75 has an extremely low [Ba/Eu] ratio of $-0.63$~dex and seems to be one of the rarer cases of a GC compatible with predominant r-process production (Fig. \ref{rs}).  An exception are the lighter n-capture elements Rb, Y, and Zr, which are more consistent with scaled solar r+s-process production. These elements, however, are associated with the weak s-process, which appears in massive  (M~$ \sim 20 \rm{M}_{\odot}$) stars on similar timescales as the r-process production from SNe II. The best fit to the production of the elements from Ba to Th in M 75 is found for (scaled solar) pure r-process enrichment, plus only $10\%$ of the (scaled solar) s-process yields. This means that only a small number of AGB stars contributed to the enrichment of the primordial cloud from which M 75 formed.

\begin{figure}[t!]
\resizebox{\hsize}{!}{\includegraphics[clip=true]{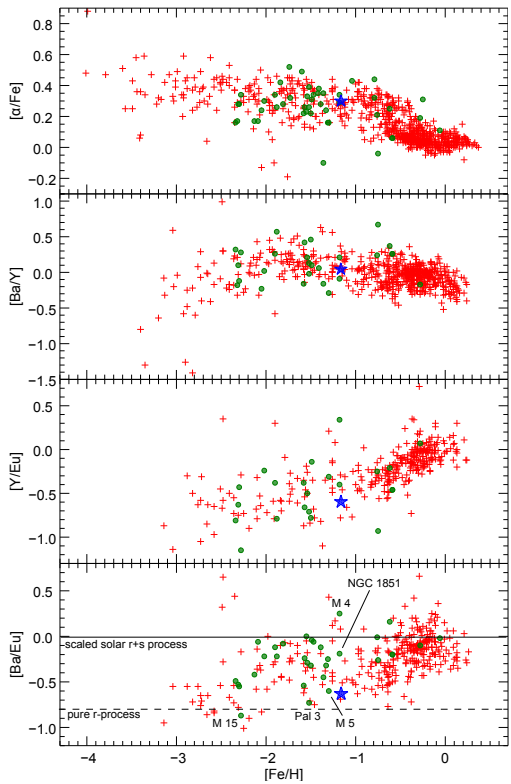}}
\caption{\footnotesize
A comparison of the mean $\alpha$ and n-capture element abundances of M 75 (blue
asterisk) with galactic disk and halo field stars (red crosses) and mean ratios of various GCs
(green circles).
}
\label{galaxy}
\end{figure}

Pure r-process enhancement is typical for metal poor halo field and GC stars with [Fe/H]~$<2.0$~dex \citep[e.g.][]{sneden+2000}, but there are also examples of GCs and halo field stars with predominant r-process enrichment at higher metallicities, e.g. NGC~3201 \citep{gonzalez+wallerstein1998} and Pal~3 \citep{koch+2009} with [Fe/H]~$-1.6$~dex. Most similar to M~75 in this respect is the GC M~5 with [Fe/H]~$=-1.3$~dex and [Ba/Eu]~$=-0.60$~dex \citep{yong+2008}, see Fig. \ref{galaxy}.

\section{M~75 and its place in the Galaxy}

In order to place M~75 amongst other GCs and halo field stars, we plotted four key abundance ratios, important to trace the chemical evolution of any stellar population. The compilation of Galactic stars is taken from \citet{venn+2004}, and the mean abundances of various GCs are from \citet{pritzl+2005}, complemented with more recent results for NGC~1851 \citep{carretta+2011}, M~5 \citep{yong+2008}, and the outermost GCs Pal~3 \citep{koch+2009} and Pal~4 \citep{koch+cote2010}. The mean (Mg, Ca, Si)-abundance is representative of M~75's $\alpha$-content, [Ba/Y] represents the main s- to weak s-process ratio, [Y/Eu] -- the weak s- to main r-process, and [Ba/Eu] -- the main s- to main r-process. The chemical abundances of M~75 are fully compatible with the bulk of Galactic GCs and halo field stars, which rules out possible extragalactic origin and accretion on a later stage to the Milky Way halo.  

\section{Conclusions}

M 75 likely hosts four chemically distinct stellar populations formed on a short timescale. The two most Na-rich stars are also Ba-enhanced, which prompts that the main polluters, which enriched the SG, also included intermediate mass AGB stars. The moderate Na-O anticorrelation and the lack of significant Mg variations are at odds with the very extended HB of M 75. The n-capture elements pattern is consistent with predominant r-process enrichment with a marginal contribution (about $10\%$ of the scaled solar yields) of s-process enriched material, typical for the most metal poor GCs and halo field stars, but not excluded at higher metallicities. The overall abundances of M~75 are consistent with the abundances of other inner and outer halo GCs and field stars, which suggests a similar origin with the bulk of Milky Way Globular clusters.

\begin{acknowledgements}
The authors acknowledge the Deutsche Forschungsgemeinschaft for funding from  Emmy-Noether grant  Ko 4161/1. NK is grateful to F. D'Antona and R.~G. Gratton for helpful discussions and to the committee headed by Father Funes, which decided to award this work with the Rome Globular Cluster Prize 2012.
\end{acknowledgements}

\bibliographystyle{aa}

\end{document}